\documentclass[a4paper,11pt]{article}
\usepackage[utf8]{inputenc}
\usepackage[english]{babel}

\usepackage{graphicx}
\usepackage[version=4]{mhchem}
\usepackage{pdfpages}

\usepackage[left=19mm,top=25mm]{geometry}
\usepackage[super,square,comma,numbers]{natbib} 
\usepackage{natmove} 

\RequirePackage{xcolor} 
\PassOptionsToPackage{dvipsnames}{xcolor}
\definecolor{Cgreen}{rgb}{0,.6,0}
\definecolor{Cblue}{rgb}{0,0,.8}
\definecolor{Clila}{rgb}{.8, 0, .8}
  \usepackage[pdftex,
              plainpages=false,
              pdfpagelabels,
			        pdfusetitle,	
              bookmarksnumbered,
              linkbordercolor={0 0 0},
              hidelinks=true,
              colorlinks={ true },
              breaklinks=true,
              urlcolor=Clila, 
              linkcolor=Cblue, 
              citecolor=Cgreen,
             ]{hyperref}

\newcommand{\Author}{Pavol Jusko,{$^{[a]}$}
                    Aude Simon,{$^{[b]}$}
                    Shreyak Banhatti,{$^{[c]}$}
                    Sandra Br\"unken,{$^{[d]}$}
                    Christine Joblin{$^{[a],*}$}}

\newcommand{\TITLE}{Direct Evidence of the Benzylium and Tropylium Cations
as the two long-lived  Isomers of \CsHs}
\title{\TITLE}
\author{\Author}

\newcommand{\rcm}{{\ensuremath{\textrm{cm}^{-1}}}}

\newcommand{\kjm}{{\ensuremath{\textrm{kJ}\cdot\textrm{mol}^{-1}}}}
\newcommand{\kmm}{{\ensuremath{\textrm{km}\cdot\textrm{mol}^{-1}}}}
\newcommand{\mjp}{{\ensuremath{\textrm{mJ}/\textrm{pulse}}}}

\newcommand{\CsHs}{{\ce{C7H7+}}}
\newcommand{\CsHe}{{\ce{C7H8}}}

\newcommand{\CeHnp}{{\ce{C8H9+}}}
\newcommand{\trp}{{\ce{Tr+}}}
\newcommand{\bzp}{{\ce{Bz+}}}
\newcommand{\Ne}{{\ce{Ne}}}
\newcommand{\ie}{{i.\,e.}}
\newcommand{\eg}{{e.\,g.}}

\begin{document}

	{\raggedright \Large\bf \TITLE\par}
	\vspace{0.5cm}
	{\raggedright \Large \Author\par}

\let\thefootnote\relax\footnotetext{$^{[a]}$~Institut de Recherche en Astrophysique et Plan\'etologie (IRAP), 
Universit\'e de Toulouse (UPS), CNRS, CNES, 9 Av. du Colonel Roche, 31028 Toulouse Cedex 4, France.}
\let\thefootnote\relax\footnotetext{$^{[b]}$~Laboratoire de Chimie et Physique Quantiques LCPQ/IRSAMC,
Universit\'e de Toulouse (UPS) and CNRS, 118 Route de Narbonne, 31062 Toulouse, France.}
\let\thefootnote\relax\footnotetext{$^{[c]}$~I. Physikalisches Institut, Universit\"at zu K\"oln, 
Z\"ulpicher Str. 77, 50937 K\"oln, Germany.} 
\let\thefootnote\relax\footnotetext{$^{[d]}$~Radboud University, Institute for Molecules and Materials, FELIX 
Laboratory, Toernooiveld 7c, 6525 ED, Nijmegen, The Netherlands.}
\let\thefootnote\relax\footnotetext{$^{*}$~christine.joblin@irap.omp.eu}

	\vspace{0.5cm}
\begin{centering}
    {Submitted version. Final version in Chem. Phys. Chem.; {\url{http://dx.doi.org/10.1002/cphc.201800744}}\par}
\end{centering}



\begin{abstract}
Disentangling the isomeric structure of \ce{C7H7+} is a longstanding experimental issue. 
We report here the full mid-infrared vibrational spectrum of \ce{C7H7+} tagged with Ne obtained
with infrared-predissociation spectroscopy at $10\;\text{K}$. 
Saturation depletion measurements were used to assign the contribution of benzylium 
and tropylium isomers
and demonstrate that no other isomer is involved.
Recorded spectral features compare well with density functional theory calculations.
This opens perspectives for a better understanding and control of the formation paths 
leading to either tropylium or benzylium ions.

%
\end{abstract}

	\vspace{0.5cm}

Benzylium (\bzp) and tropylium (\trp) ions are key isomers of \CsHs\ commonly produced from energized 
toluene (\CsHe) \cite{Dunbar1975, Kuck1990, Lifshitz1994}.
Whereas \bzp consists of a benzene ring substituted with a methylene group, the  \trp\ 
isomer is a fully aromatic ion made of a 7-membered CH ring. The possibility that similar 
structures can be involved in the fragmentation of methyl-substitued polycyclic aromatic 
hydrocarbon ions has been recently discussed \cite{Rapacioli2015, Jusko2018}.
Still, one of the major issues in these studies is the limited understanding of the 
formation paths of these species, that would allow us to predict and therefore
control the production of one or the other isomer.
On the one hand, it was shown that the choice of the precursor is important, \eg,
the use of halogen-substitued toluene is expected to optimise 
the production of \bzp \cite{Jackson1977}. But on the other hand,
the \bzp/\trp\ ratio depends strongly on the 
various experimental conditions \cite{Fridgen2004,Morsa2014}. 
Another point, in photoionization experiments of toluene, is that it is not possible to 
produce solely 
\trp\ at appearance threshold  \cite{Lifshitz1993jpc}, despite calculations show
 \trp\ being energetically more favorable than 
\bzp\ (by $\sim38\;\kjm$; cf. Tab~S3 in the SI).
The latter results were interpreted as due both to the presence of a barrier in the dissociation 
path from toluene towards \trp, and to the role of autoionizing states in promoting a 
nonstatistical formation of \bzp\ over \trp\ \cite{Lifshitz1993jpc}.
In contrast to \trp, theoretical studies show that paths towards \bzp\ lack high barriers, 
and no barrier is found in the path from benzyl chloride  ionization towards \bzp\ 
formation \cite{Choe2008}.

It is known since earlier studies on \CsHs\ that a convenient way to identify \bzp\ is
through ion-molecule reaction with toluene, leading to efficient formation  of 
\CeHnp \cite{Shen1974,Dunbar1975}.
The non-reactive part of the ion population is then attributed to \trp.
In parallel, there have been many attempts to characterize the structure of 
\CsHs\ directly, \ie, via spectroscopy.
Electronic absorption features have been studied in both {Ne} matrix \cite{Nagy2011} 
and in gas-phase by photodissociation \cite{Dryza2012,Feraud2014}.
For  \bzp, the vibronic structure was resolved for the S$_1$$\leftarrow$S$_0$ 
transition but much broader features were observed for the higher excited states.
Absorption features of \trp\ are expected in the UV range \cite{Nagy2011}, but the 
data obtained in Ne matrices could not be confirmed in the gas-phase  \cite{Dryza2012}.
On the opposite, infrared and Raman spectra of \trp could be obtained in solutions or 
in solid phase using salts \cite{Fateley1957,Sourisseau1978,Howard1985}.
In gas-phase, one can think of infrared multiple photon dissociation (IRMPD) vibrational 
spectroscopy as the technique of 
choice when characterizing the structure of ions in mass spectrometry 
experiments \cite{Oomens2006}.
Information on the \bzp/ \trp\ dichotomy  \cite{Chiavarino2012} could be obtained from 
the IRMPD spectra of derivative ions with lower dissociation thresholds compared 
to \CsHs \cite{Chiavarino2006, Schroeder2006, Chiavarino2012, Zins2010}.
The spectroscopy of polycyclic hydrocarbon cations containing \bzp\ and \trp\ 
structure was also investigated \cite{Morsa2014, Jusko2018}.
However, the IRMPD spectra of pure \bzp\ and \trp\ are still lacking, despite the efforts \cite{Chiavarino2006}.
This could be due to the exceptional stability of these ions, or
possible isomerisation issues, as found in the case of \ce{C17H11+} \cite{Jusko2018}.

\begin{figure}[!h]
\centering
  \includegraphics[width=0.5\linewidth]{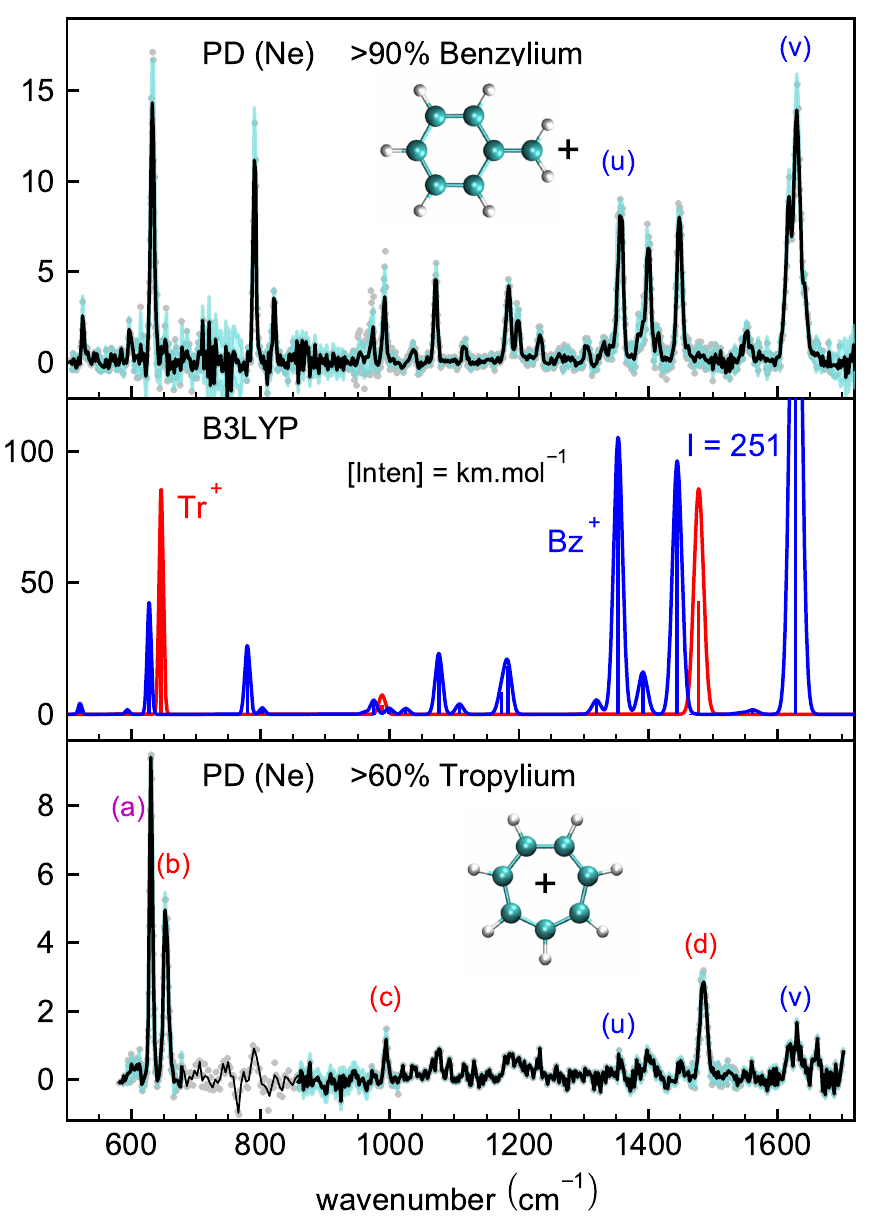}
  \caption{
  Mid-IR spectrum of \CsHs.
  Top panel -- IR pre-dissociation spectrum of \bzp\ tagged with Ne.
  Middle panel -- stick spectrum corresponding to the scaled DFT  harmonic spectra 
  for \bzp\ (blue) and \trp\ (red).
  Convoluted spectra with $\sigma=0.5\%$ BW are provided for comparison with the experiment.
  Bottom panel -- IR pre-dissociation spectrum of \trp\ tagged with Ne.
  Bands marked by (a), (b), (c), (d) have been confirmed to belong to
  \trp\  and those marked (a), (u), (v) to \bzp\ by saturation depletion measurements.
  Intensities for experimental spectra are in 
  arbitrary units. All recorded bands are listed in Tab.~\ref{t:bzp} for \bzp\ and 
  Tab.~\ref{t:trp} for \trp.
    }\label{f:exp}
\end{figure}

\begin{figure}[!htb]
\centering
 \includegraphics[width=0.5\linewidth]{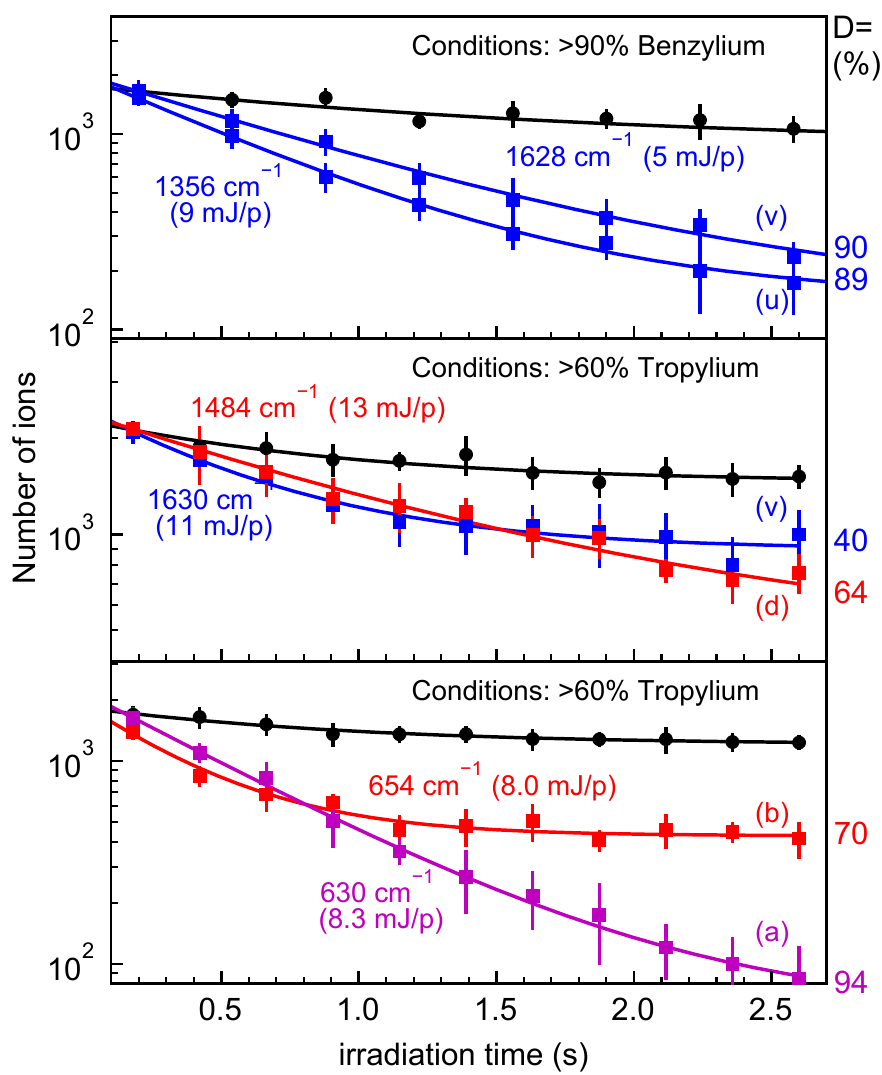}
\caption{Depletion of $\CsHs\cdot\Ne$ in the ion trap for the different experimental conditions used in Fig.1.
Color code:  Blue -- wavelength at which the \bzp\ isomer is active.
Red -- wavelength at which the \trp\ isomer is active. Magenta -- both isomers active.
Black -- corresponds to non-resonant dissociation of 
the Ne tag ($\sim 8-10\;\mjp$).
Numbers in the right column correspond to saturation depletion in percent. 
Letters refer to the bands shown in
Fig.~\ref{f:exp}.
}\label{f:depl}
\end{figure}

An alternative approach to IRMPD 
not owning the disadvantages of the multi-photon process, is infrared pre-dissociation 
(IR-PD) spectroscopy of a weakly bound complex of the ion with a rare gas atom.
Thanks to this technique, we were able to obtain the first complete mid-IR-PD 
spectra of \Ne\ tagged \CsHs\ recorded in gas-phase and at low temperature.
We used the cryogenic 22 pole ion trap \cite{Asvany2010}, which is coupled to the 
free electron laser FELIX, as described in the experimental section below.
We introduced different precursors and varied the 
ionization conditions (cf. Tab.~S1 in the SI). 
Figure~\ref{f:exp} gathers the recorded experimental spectra, which appear to agree 
well with the calculated spectra of \bzp\ and \trp\ using density functional theory (DFT)
at the B3LYP/6-31G(d,p) level of theory as implemented in Gaussian09 suite of 
programs \cite{Gaussian2009}.
In addition, the contribution of \bzp\ and \trp\  in each mixture could be determined 
directly during the spectroscopic experiment by using saturation depletion measurements.
To achieve the latter, the laser is tuned in resonance with a vibrational band of interest. 
By applying a large number of photon pulses, the optically active isomer-Ne complex 
can be dissociated completely leaving only the optically inactive isomer-Ne complex 
in the trap. 
Recording the ion-Ne ($m=111\;u$) number as a function of time reveals the relative 
abundance of active to inactive isomers. Depletion tests for selected bands of both 
isomers and under varying ion source conditions are shown in Fig.~\ref{f:depl}.
They show that one condition leads to 90\% \bzp\ population and another to
60-70\% of \trp\ with 40-30\% of \bzp.
These measurements provide the first complete demonstration that  \bzp\  and  \trp\ are the only 
long-lived isomers of \CsHs.
In a preparatory study, we also recorded the chemical reactivity with toluene and 
found a good agreement with the above measurements, although only \bzp\ can be 
directly monitored (cf. Fig.~S2 in the SI).

 
Assignment of the experimental bands of \bzp\ and \trp\ is made in 
Tables~\ref{t:bzp} and~\ref{t:trp}. 
Additional levels of theory were carried out for benchmarking (cf. Section 2 in the SI).
For each level, the calculated band positions are linearly scaled with a scaling factor S, 
in order to match with the strongest \bzp\ band.
From calculations on \Ne\ tagged ions, we found that the 
top (T) and molecular plane (P) isomers are quasi degenerate (cf. Tab.~S3 in the SI) and that the Ne tag is 
not expected to induce band shifts (cf. Figs.~S12 and S13, Table~S4 in the SI), nor
splitting of the bands due to the degeneracy lifting in the \trp\ case, that would be 
observable in our experiments (cf. Fig.~S12 in the SI, degenerate bands always 
split by less than $1\;\rcm$).
Overall, we found a good agreement between the measured and calculated band positions.
However, in the case of \trp, one major difference exists. 
The harmonic calculations provide only three infrared active bands, which can be 
rationalized by the high symmetry of \trp\ which
 pertains to the ${\rm D}_{\rm{7h}}$ point group symmetry 
(cf. Tab.~S2 in the SI for descent in symmetry from ${\rm D}_{\rm{7h}}$ to C$_{2v}$, 
in which calculations are performed).
In the experimental spectrum, we found four bands,
two bands including the additional one falling at low frequency.
We note that two close-by bands in this range are also observed
in some experiments in condensed phase \cite{Fateley1957, Sourisseau1978, Howard1985}.

\begin{table}[!h]
    \caption[]{
    Experimental mid-IR band positions recorded for \bzp\ and calculated values using 
    DFT (B3LYP/6-31G(d,p)) obtained in the present work (see Figure~\ref{f:exp}). 
    Only the modes which are IR active are reported.
    Their symmetry in the point group of the molecular ion (C$_{2v}$) are reported.
    }\label{t:bzp}
    \begin{center}
    \begin{tabular}{ccccc}
        \hline
        Mode & Exp. & \multicolumn{3}{c}{Calc.} \\
          & $\nu$ & $\nu$ & Inten.& Sym. \\
        \hline
        $\nu_{5}$   &   524    &      521  &	    4.3 &A$_1$\\
        $\nu_{6}$   &   596    &      594  &	    1.8 &B$_2$  \\
        $\nu_{8}$   &   632    &      628  &	   42.8 &B$_1$  \\
        $\nu_{9}$   &   790    &      780  &	   26.3 & B$_1$ \\
        $\nu_{10}$  &   820    &      803  &	    2.6 & A$_1$\\
        $\nu_{12}$  &          &      962  &	    0.9 & B$_2$  \\
        $\nu_{13}$  &   974    &      976  &	    5.4 & A$_1$ \\
        $\nu_{16}$  &   992    &     1000  &	    2.4 & A$_1$ \\
        $\nu_{17}$  &   1036   &     1026  &	    2.3 & B$_1$\\
        $\nu_{18}$  &   1070   &     1077  &	   23.2 & B$_1$ \\
        $\nu_{19}$  &   1114   &     1109  &	    4.0 & B$_2$ \\
        $\nu_{20}$  &   1184   &     1174  &	    8.4 & B$_2$\\
        $\nu_{21}$  &   1198   &     1184  &	   18.5 & A$_1$  \\
                    &   1232   &           &	        \\
                    &   1302   &           &	        \\
        $\nu_{22}$  &   1332   &     1321  &	    5.6 & B$_2$\\
        $\nu_{23}$  &   1356   &     1355  &	  105.5 & A$_1$ \\
        $\nu_{24}$  &   1400   &     1393  &	   16.1 & B$_2$  \\
        $\nu_{25}$  &   1448   &     1446  &	   96.4 & B$_2$ \\
        $\nu_{26}$  &          &     1467  &	    0.2 & A$_1$\\
        $\nu_{27}$  &          &     1541  &	    0.6 & B$_2$ \\
        $\nu_{28}$  &   1554   &     1564  &	    1.8 & A$_1$\\
        $\nu_{29}*$ &   1630   &     1630  &	  250.9 & A$_1$\\
        \hline
    \end{tabular}
    \vskip 0.2em
    \end{center}
    {\footnotesize{\emph{Note:}
    All frequencies in \rcm, intensities in \kmm.
    * -- mode used to determine the scaling 
    factor 0.974. Cf. Fig.~S15. in the SI for the visualisation of modes.
    }}
\end{table}

\begin{table}[!h]
    \caption[]{
    Experimental mid-IR band positions recorded for \trp\ and calculated values using
    DFT (B3LYP/6-31G(d,p)) obtained in the present work (see Figure~\ref{f:exp}). 
    Only the modes which are IR active  are reported. 
    Although they were computed in the C$_{2v}$ symmetry point group, their symmetry in the 
    point group of the molecular ion (D$_{7h}$) are listed.
    }\label{t:trp}
    \begin{center}
    \begin{tabular}{ccccc}
        \hline
        Mode & Exp.     & Ref.  &\multicolumn{2}{c}{Calc.} \\
             & $\nu$    & $\nu$ & $\nu$ (Inten.)   & Sym. \\
        \hline
         $\nu_{\rm{c}}$ &  630   & 633\cite{Fateley1957}, 634\cite{Sourisseau1978}, 655\cite{Howard1985} &  
                        &           \\
        $\nu_{4}$       &  652   & 658\cite{Fateley1957}, 660\cite{Sourisseau1978}, 685\cite{Howard1985} &
         646(85.8)	    & A''$_{2}$ \\ 
        $\nu_{8}$       &  994   & 992\cite{Fateley1957}, 992\cite{Sourisseau1978}   & 989(3.7)   & E'$_{1}$ \\
        $\nu_{14}$      &  1486  & 1477\cite{Fateley1957}, 1477\cite{Sourisseau1978} & 1480(42.9) & E'$_{1}$ \\
        \hline
    \end{tabular}
    \end{center}
    \vskip 0.2em
    {\footnotesize{\emph{Note:}
    All frequencies in \rcm, intensities in \kmm. Scaling factor 0.974.
    $\nu_{\rm{c}}$ is a combination band of the two lowest lying fundamental bands of \trp.
    Cf. Fig.~S16 -- S18. in the SI for the visualisation of modes. 
    }}
\end{table}

Saturation depletion measurements established that bands (b) -- (d) in the experimental 
spectrum (Fig.~\ref{f:exp}), 
belong solely to \trp, and that both \bzp\ and \trp\ are
active at the $630\;\rcm$ band (a).
Bands (c) and (d) can be assigned to the doubly degenerate in plane CH bending mode 
and in plane CC stretching mode of \trp, respectively.
The third \trp\ band, the CH out-of-plane bending mode (not degenerate) should lay around 
$650\;\rcm$ (close to band (b), cf. Fig.~S3).
The performed saturation depletion measurements invalidate the hypothesis that
both bands (a) and (b) are caused by distinct (T) and (P) $\trp\cdot\Ne$ isomers.
We also excluded contribution from a triplet spin state, whose presence can 
lead to shifts in low energy modes for PAHs \cite{Falvo2012}. In the case of \trp, it was 
found distorted and $3.3\;\text{eV}$ above the ground state at the 
B3LYP/6-31G(d,p) level of theory (cf. Fig.~S14 in the SI for its computed IR spectrum). 
The excited state can not survive many collisions with cold buffer gas and its 
spectrum does not contain features in the wavelength range of interest.
The most likely scenario to account for bands (a) and (b), 
is that a combination band comes into close resonance with the fundamental 
CH out-of-plane 
bending mode, and borrows a significant part of its intensity.
Our calculations show that the combination band involves the
two lowest vibrational modes of \trp\ and that the combination band is red shifted
relative to the fundamental band (Tab.~\ref{t:trp} and Tab.~S5 in the SI).
In our experiment, we can assign band (b) as the fundamental mode (\ie\ the strongest band) 
since the initial dissociation rate of this band is higher and its band width at saturation 
is larger as compared to band (a) (Fig.~\ref{f:depl}).
Finally, our spectra do not include the CH stretch range. However, we learned while writing
this manuscript that this range has recently been studied 
using a jet experiment and a table-top laser \cite{DUNCAN}.
%
%
%

We recorded the mid-IR-PD spectra of \CsHs\ tagged with Ne at $10\;\text{K}$.
The comparison of the experimental spectra with the harmonic calculated spectra 
show, that neither significant band shifts, nor band splitting are induced by
the presence of the Ne tag, that can in particular induce a symmetry decrease 
in the case of \trp.
The \trp\ spectrum exhibits an interesting strong resonance at low frequency 
between a combination band and the fundamental CH out-of-plane bending mode. 
Anharmonic calculations would help 
in providing a detailed quantitative study of this resonance \cite{Mackie2015, Mulas2018}. 

We used saturation depletion measurements to confirm the existence of only two long-lived 
isomers of \CsHs,  and the obtained spectra support their initial structural assumption  
as \bzp\ and \trp\ \cite{Dunbar1975, Lifshitz1993jpc}.
The depletion technique in an ion trap has the advantage of being able to track the abundance
of a well-defined isomer, contrary to the earlier reactivity method, which only      has access to \bzp.
This technique opens new perspectives for our understanding of the isomerisation paths of \CsHs.


\section*{Experimental Section}

The ions are produced by electron bombardment in a storage type ion source 
\cite{Gerlich1992} from benzyl chloride and toluene precursors. 
Ions of mass $91\;u$ (\CsHs) are mass selected in a quadrupole mass filter prior to being 
injected into the cryogenic 22 pole rf ion trap (nominal temperature $8-9\;\rm{K}$), 
where they are cooled by a $100\;\rm{ms}$ long 3:1 He:Ne  gas pulse. The high 
number gas density achieved during this pulse  ($n\sim10^{14}\;{\rm{cm}}^{-3}$) promotes 
ternary attachment of \ce{Ne} to the bare ion producing 
$\bzp\cdot\Ne$ and $\trp\cdot\Ne$ complexes. 
After the cooling and tagging sequence, the number density in the trap decreases 
quickly to a level where no further attachment takes place. The ion--Ne complex is  
then dissociated by infrared radiation provided by the FELIX FEL-2 free electron 
laser \cite{Oepts1995}, operated in the $500-1650\;\rcm$ range.
The laser delivers up to $30~\text{mJ}$ in a single macropulse into the 22 pole 
trap with a repetition rate of $10\;\rm{Hz}$ and spectral bandwidth better 
than $\sigma=0.5\%$. After an irradiation time of typically $2.6\;\rm{s}$ the trap 
content is emptied through a mass filter onto a counting detector.

\section*{Acknowledgements}

The research leading to this result is supported by the European Research 
Council under the European Union's Seventh Framework Programme ERC-2013-SyG, 
Grant Agreement n. 610256 NANOCOSMOS. We acknowledge support from the project
CALIPSOplus under the Grant Agreement 730872 from the EU Framework
Programme for Research and Innovation HORIZON 2020. 
We greatly appreciate the experimental support provided by the FELIX team.
We gratefully acknowledge the Nederlandse Organisatie voor Wetenschappelijk 
Onderzoek (NWO) for the support of the FELIX Laboratory.
We thank the Cologne Laboratory Astrophysics group
for providing the FELion ion trap instrument
and the Cologne Center for Terahertz Spectroscopy (core facility,
DFG grant SCHL 341/15-1) for supporting its operation.
S.Ba. is supported by the H2020-MSCA-ITN-2016 Program (EUROPAH project, G. A. 722346).
A. S. thanks the computing facility CALMIP for generous 
allocation of computer resources.

\vspace{1 cm}
\noindent {\bf Keywords:} benzylium; tropylium; structure elucidation; IR spectroscopy; 
cryogenic ion trap


\bibliography{bztr}

\clearpage

\section*{Entry for the Table of Contents}

\includegraphics{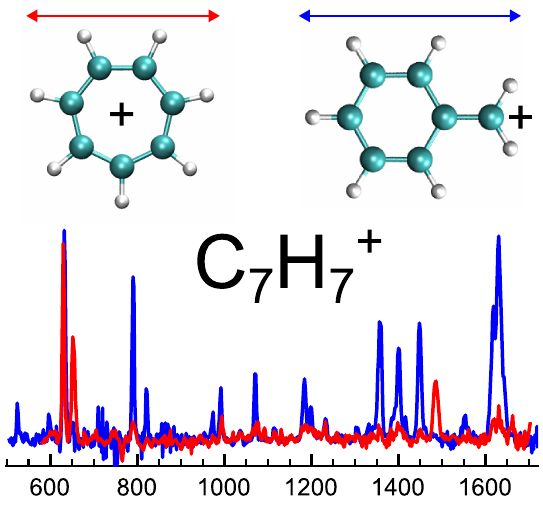}

\vspace{1 cm}

{\bf \CsHs\ isomers:} by using infrared pre-dissociation spectroscopy and saturation 
depletion measurements of $\CsHs\cdot\Ne$ complexes,
we demonstrate that benzylium and tropylium cations are the only two long-lived 
isomers of gas-phase \CsHs.
\clearpage



\section*{Supplementary material}

# Supplementary Information

# Direct Evidence of the Benzylium and Tropylium Cations as the two long-lived Isomers of $C_7H_7^+$

Pavol Jusko,[a] Aude Simon,[b] Shreyak Banhatti,[c] Sandra Brünken,[d] Christine Joblin[a],*

July 31, 2018

---

[a] Institut de Recherche en Astrophysique et Planétologie (IRAP), Université de Toulouse (UPS), CNRS, CNES, 9 Av. du Colonel Roche, 31028 Toulouse Cedex 4, France.
[b] Laboratoire de Chimie et Physique Quantiques LCPQ/IRSAMC, Université de Toulouse (UPS) and CNRS, 118 Route de Narbonne, 31062 Toulouse, France.
[c] I. Physikalisches Institut, Universität zu Köln, Zülpicher Str. 77, 50937 Köln, Germany.
[d] Radboud University, Institute for Molecules and Materials, FELIX Laboratory, Toernooiveld 7c, 6525 ED, Nijmegen, The Netherlands.
* christine.joblin@irap.omp.eu

# Contents

# 1 Experimental

## 1.1 Mass resolution of the experimental setup

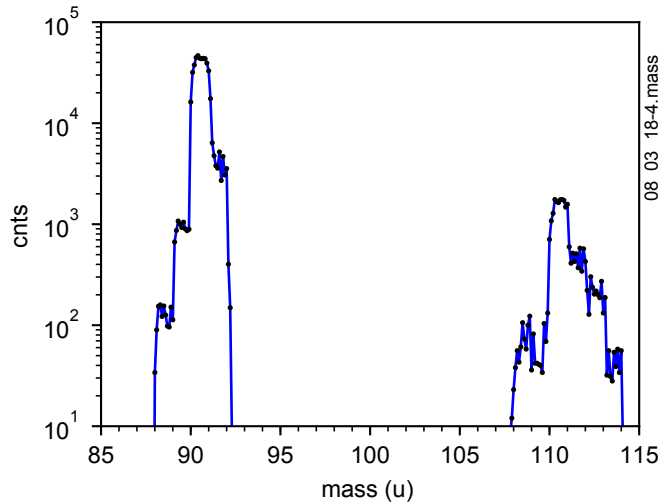

Figure S1: Mass resolution of the experimental setup. Ions of interest ($C_7H_7^+$, 91 $u$) are mass selected in the source quadrupole and injected into the trap. At the same time, a buffer gas pulse (He:Ne mixture cca. 3:1, pulse length 100 ms) is applied in order to aid thermalisation and promote ternary attachment. The ionic products are extracted after 500 ms storage time and mass selected in the product quadrupole (abscissa). Nominal trap temperature is 9 K. We achieved up to 5 % Ne attachment.

## 1.2 Ion production

Table S1: Precursors and source conditions used for ion production. Chemical probing (cf. Section 1.3) is used to obtain the abundance of benzylium.

| Precursor | Benzyl chloride $C_7H_7Cl$ | Toluene $C_7H_8$ | Toluene $C_7H_8$ |
|---|---|---|---|
| Electron energy (approx.) (eV) | 25 | 11* | 25 |
| Benzylium$^+$ (%) | $> 90$ | $\sim 30$ | $\gtrsim 60$ |

**Note:** Ions are produced in the storage ion source. They are extracted in a pulse shorter than 1 ms, after residence exceeding tens of ms. The amount of Bz$^+$ we report in this table, is not strictly the electron bombardment yield, but rather the product of ion-molecule reactions inside the source.

*– lowest electron energy for reasonable ion yield.

## 1.3 Isomer composition – chemistry

The reaction of $C_7H_7^+$ with toluene

$$Bz^+ + C_7H_8 \xrightarrow[reaction]{fast} C_8H_9^+ + C_6H_6, \qquad (1)$$

$$Tr^+ + C_7H_8 \xrightarrow[reaction]{no} \qquad , \qquad (2)$$

is used to determine the fraction of $Bz^+$ in the mixture. Time evolution of reactants and products in the trap is shown in Fig. S2.

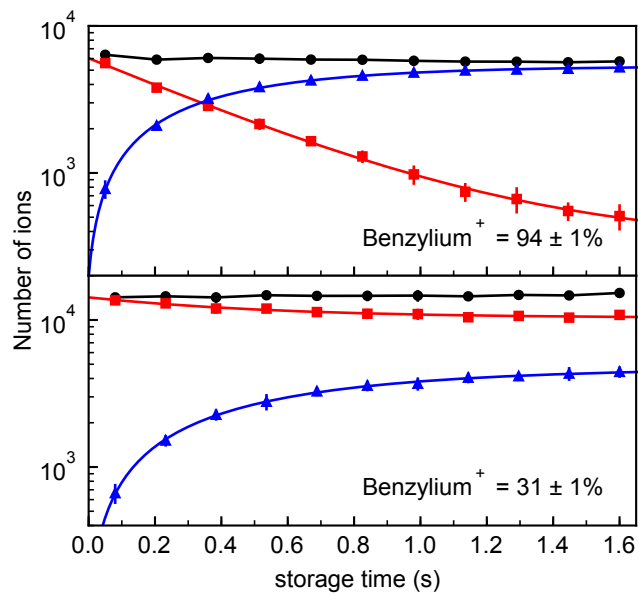

Figure S2: Number of ions in the trap as a function of the storage time. Red – $C_7H_7^+$ (91 $u$). Blue – $C_8H_9^+$ (105 $u$). Black – sum af all ions. Full lines represent exponential increase/ decay fits. The ratio between the reactive and non-reactive component of $C_7H_7^+$ is calculated from the fitted parameters. We show the best conditions achieved for $Bz^+$ isomer (upper panel) and $Tr^+$ isomer (lower panel).

Note: Ions are trapped using a He pulse (tens of ms) at the beginning of the cycle. Neutral reactant, toluene, is leaked into the trap continuously mixed with He carrier gas ($\sim 3.5\%$) and its estimated number density is $8 \cdot 10^8$ cm$^{-3}$. Nominal trap temperature is $240 - 250$ K.

# 2 Calculations

All calculations were performed in Gaussian09. They were computed in the $C_{2v}$ symmetry point group. The following methods and basis sets were used:

- B3LYP/6-31G(d,p)
- B3LYP/cc-pvdz
- B3LYP/cc-pvtz
- wB97XD/cc-pvdz
- wB97XD/cc-pvtz
- MP2/cc-pvtz
- B2PLYP/cc-pvtz

Molecules $Bz^+$, $Bz^+ \cdot Ne(T, S)$, $Tr^+ \cdot Ne(S)$ belong to $C_{2v}$ point group. Molecules $Tr^+$ and $Tr^+ \cdot Ne(T)$ belong to $D_{7h}$ point group. The descent in symmetry from $D_{7h}$ point group to $C_{2v}$ point group is presented in Tab. S2.

Table S2: The descent in symmetry from $D_{7h}$ to $C_{2v}$ for the $Tr^+$ ion (experimentally observed bands only). Calculated values ($\nu$, Inten.) correspond to the B3LYP/6-31G(d,p) level of theory.

| Mode | Exp. | Calc. | | | |
|---|---|---|---|---|---|
| | $\nu$ | $\nu$ | Inten. | Sym. | |
| | (cm$^{-1}$) | (cm$^{-1}$) | (km · mol$^{-1}$) | $D_{7h}$ | $C_{2v}$ |
| $\nu_c$ | 630 | | | | |
| $\nu_4$ | 652 | 646 | 85.8 | A"$_2$ | B$_2$ |
| $\nu_8$ | 994 | 989 | 3.7 | E'$_1$ | A$_1$ + B$_1$(*) |
| $\nu_{14}$ | 1486 | 1480 | 42.9 | E'$_1$ | A$_1$ + B$_1$(*) |

*Note:* (*)– Split degeneracies. Scaling factor 0.974.

## 2.1 Molecular structure (energetics)

Table S3: Relative energies (including zero point energy corrections) of $Tr^+$ vs. $Bz^+$ (black) and $Tr^+ \cdot Ne$ (T) vs. (P) (red) and $Bz^+ \cdot Ne$ (T) vs. (P) (blue) using the different theoretical methods.

| $\Delta(E + ZPE)$ (kJ·mol$^{-1}$) | $Bz^+$ | $Tr^+$ | $Bz^+ \cdot Ne$ (T) | $Bz^+ \cdot Ne$ (P) | $Tr^+ \cdot Ne$ (T) | $Tr^+ \cdot Ne$ (P) |
|---|---|---|---|---|---|---|
| B3LYP/6-31G(d,p) | +38 | 0 | +8.6 | 0 | +7.1 | 0 |
| B3LYP/cc-pvtz | +36 | 0 | +1.9 | 0 | +1.5 | 0 |
| wB97XD/cc-pvdz | +36 | 0 | +5.7 | 0 | +5.0 | 0 |
| wB97XD/cc-pvtz | +34 | 0 | +0.4 | 0 | 0 | +0.65 |
| MP2/cc-pvtz | +42 | 0 | +0.2 | 0 | +0.07 | 0 |
| B2PLYP/cc-pvtz | +31 | 0 | +1.5 | 0 | +1.2 | 0 |

*Note:* The wB97XD/cc-pvtz and MP2/cc-pvtz levels of theory are expected to be the best approaches used to describe the systems studied in this work (wB97XD takes into account dispersion and long range interactions, MP2 is post Hartree Fock, using large cc-pvtz basis set).

## 2.2 Molecular structure of the bare ions

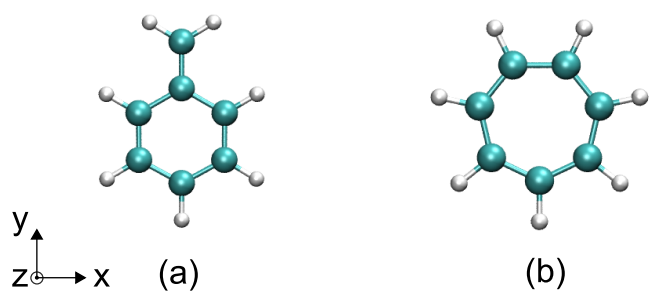

Figure S3: Structure of $Bz^+$ (a) and $Tr^+$ (b). The $Tr^+$ structure is energetically more favorable (Tab. S3).

## 2.3 Molecular structure of the ion-Ne complex

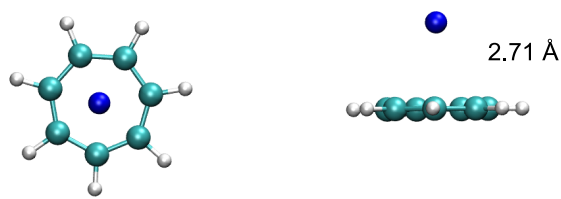

Figure S4: Structure of $Tr^+ \cdot Ne$(T) with Ne on top of the molecular plane. This structure and the structure with Ne in the molecular plane (Fig. S5) are quasi degenerate (Tab. S3, especially at the wB97XD/cc-pvtz and MP2/cc-pvtz levels of theory).

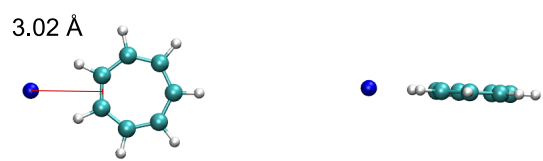

Figure S5: Structure of Tr$^+ \cdot$Ne(P) with Ne in plane with the molecule.

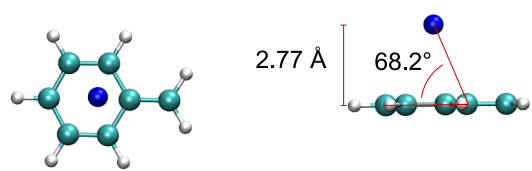

Figure S6: Structure of Bz$^+ \cdot$Ne(T) with Ne on top of the molecular plane. Calculated relative energies at the wB97XD/cc-pvtz and MP2/cc-pvtz levels of theory (Tab. S3) are well below 1 kJ$\cdot$mol$^{-1}$ for all the structures in Figs. S6 – S8, thus, we consider these as quasi degenerate.

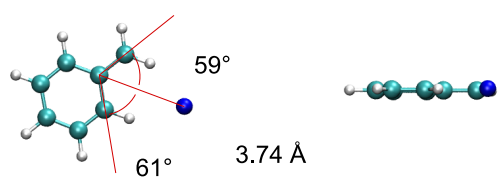

Figure S7: Structure of Bz$^+ \cdot$Ne(P) with Ne in plane with the molecule (Ne next to CH$_2$). Note that according to the calculations (Tab. S3), this is the Bz$^+ \cdot$Ne structure with the lowest energy.

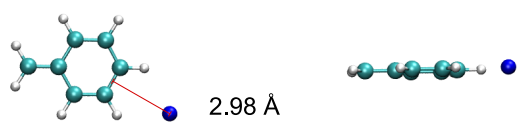

Figure S8: Structure of Bz$^+ \cdot$Ne with Ne in plane with the molecule. (Ne opposite to CH$_2$).

## 2.4 Comparison of different methods

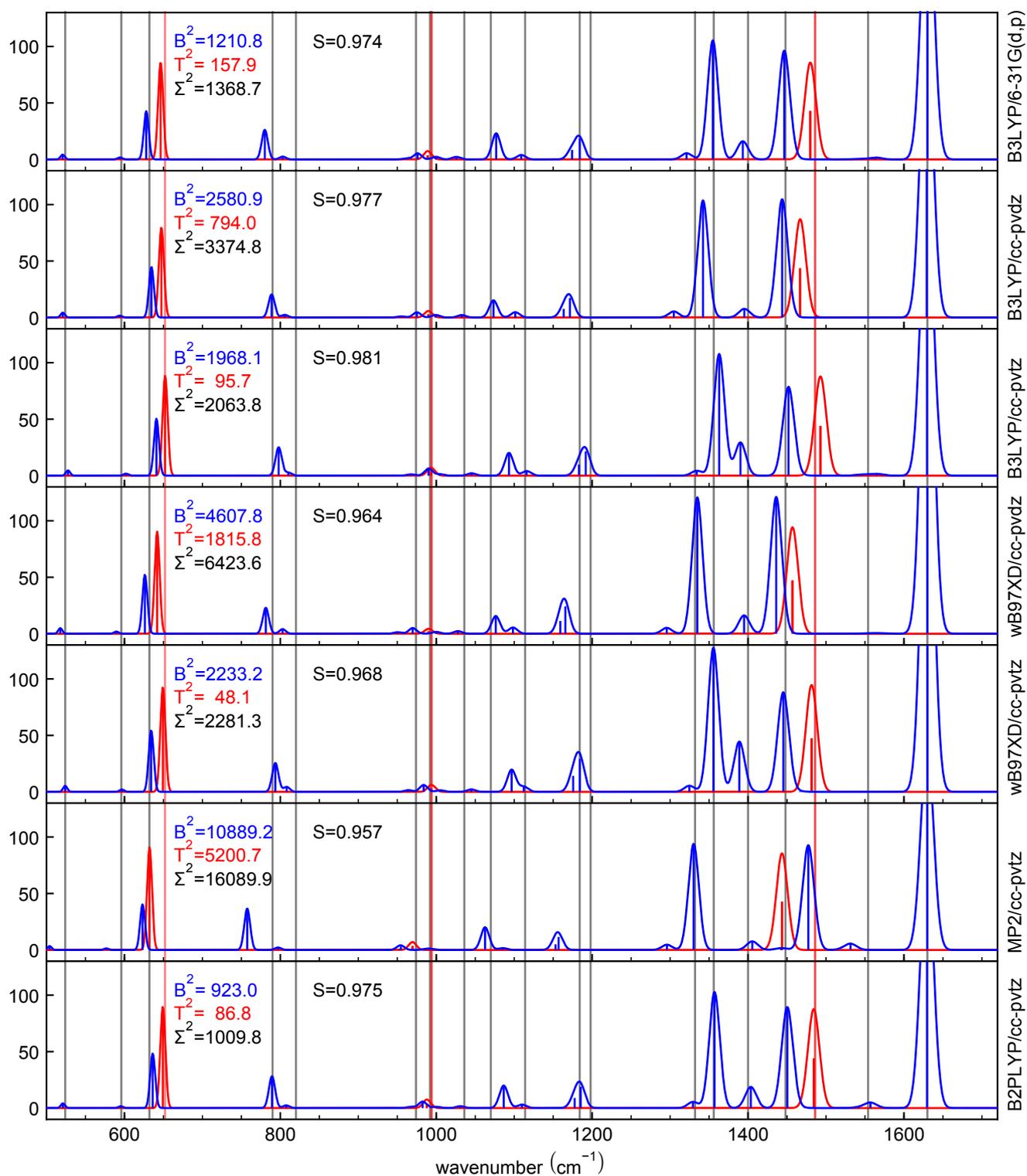

Figure S9: Comparison of the calculated spectra of Bz$^+$ and Tr$^+$, obtained by the different methods with experimental data. B$^2$, T$^2$, $\Sigma^2$ represent the sum of squares of "calculated – measured" positions for Bz$^+$, Tr$^+$, and their sum, respectively. Only bands marked with vertical lines (experimental data, Tab. 1 and Tab. 2 in the text) are used in the calculations. S corresponds to the scaling factor determined using the 1630 cm$^{-1}$ band of Bz$^+$.

## 2.5 Influence of the Ne tag

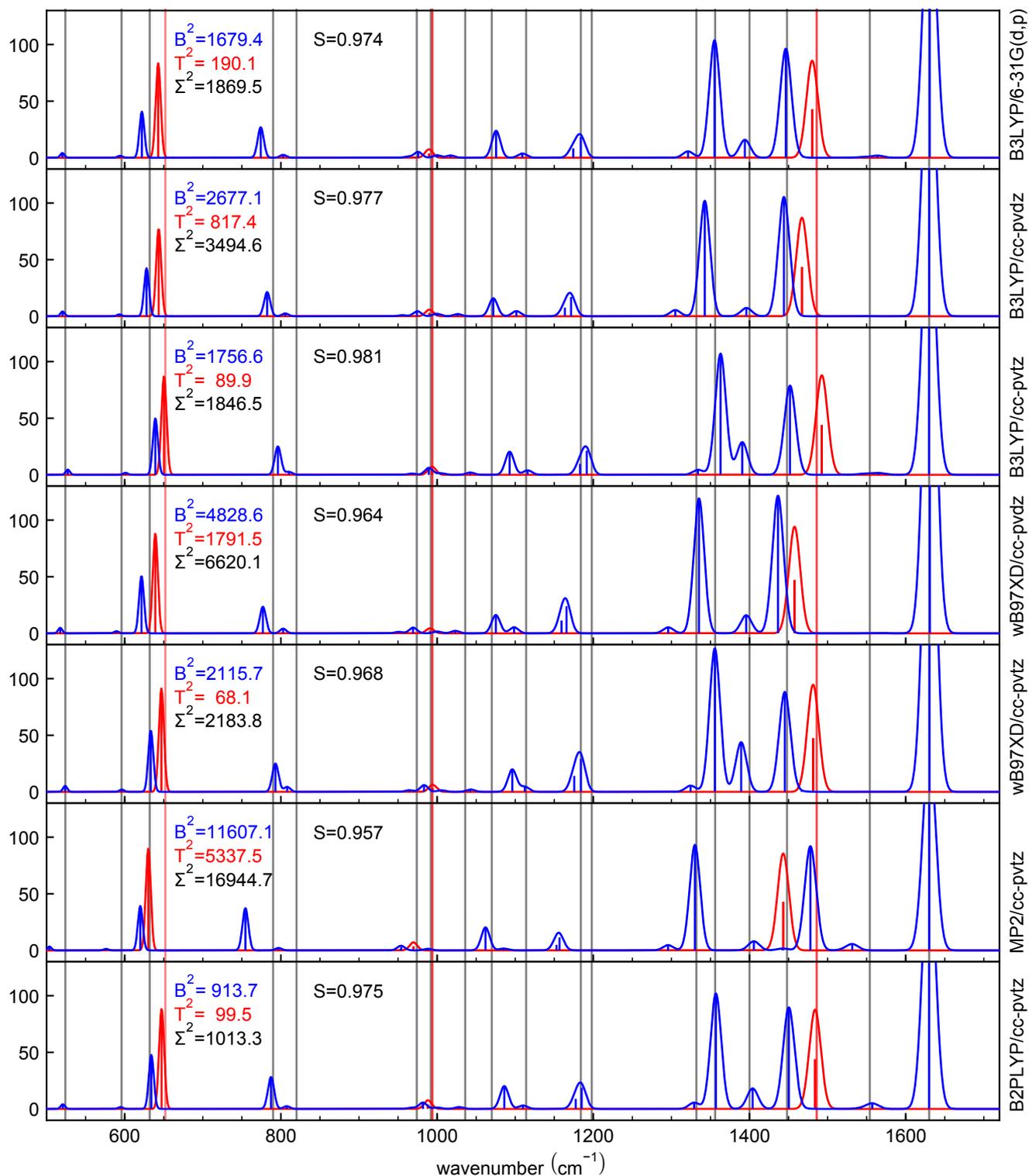

Figure S10: Comparison of the calculated spectra of $Bz^+ \cdot Ne$ and $Tr^+ \cdot Ne$, obtained by the different methods with experimental data. Ne attached on top of the molecular plane (cf. Figs. S4, S6). $B^2$, $T^2$, $\Sigma^2$ represent the sum of squares of "calculated – measured" positions for $Bz^+ \cdot Ne$, $Tr^+ \cdot Ne$, and their sum, respectively. Only bands marked with vertical lines (experimental data, Tab. 1 and Tab. 2 in the text) are used in the calculations. S corresponds to the scaling factor (determined using the 1630 cm$^{-1}$ band of $Bz^+$, cf. Fig. S9).

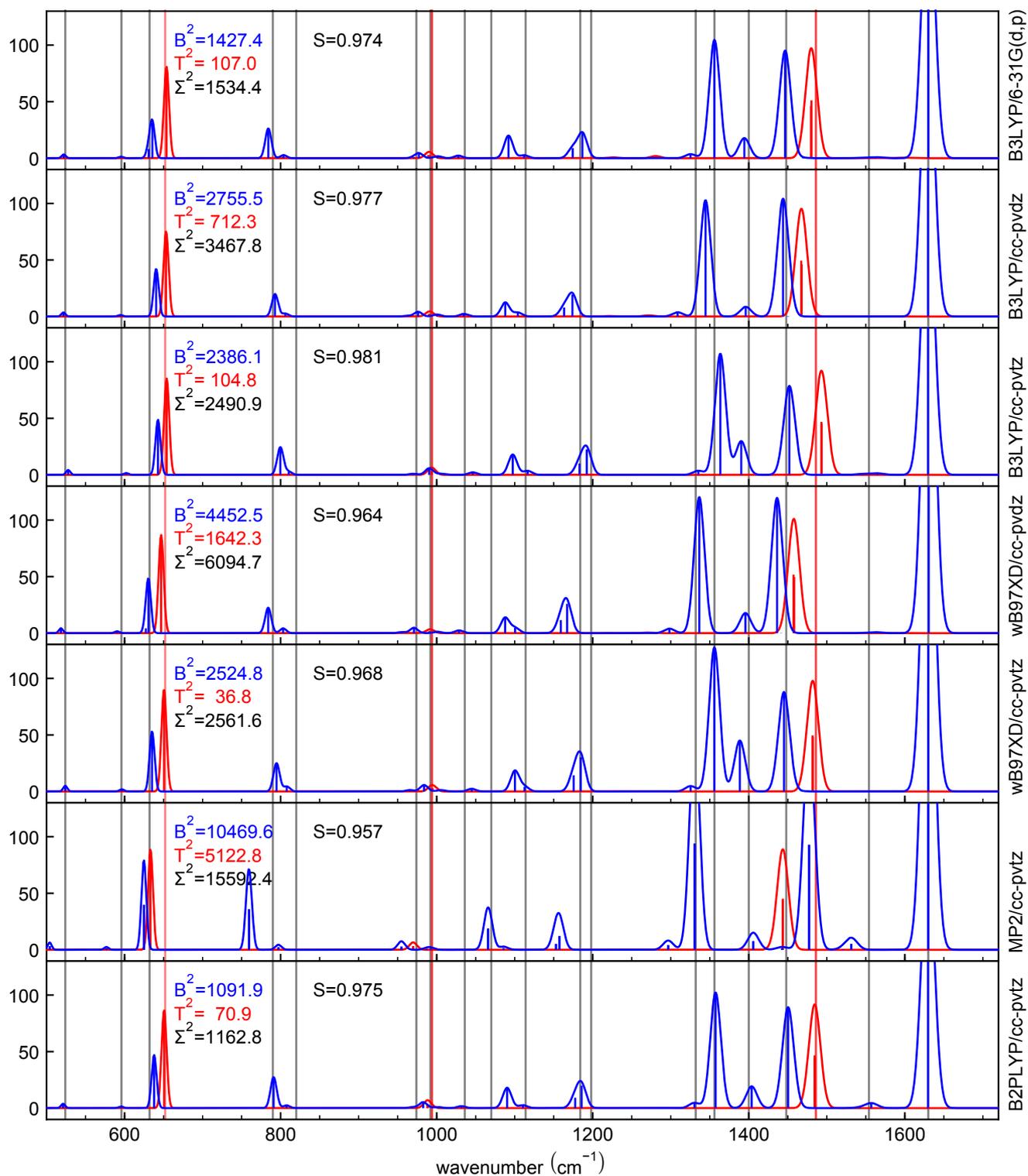

Figure S11: Comparison of the calculated spectra of $Bz^+ \cdot Ne$ and $Tr^+ \cdot Ne$, obtained by the different methods with experimental data. Ne attached in the molecular plane (cf. Figs. S5, S7). $B^2$, $T^2$, $\Sigma^2$ represent the sum of squares of "calculated – measured" positions for $Bz^+ \cdot Ne$, $Tr^+ \cdot Ne$, and their sum, respectively. Only bands marked with vertical lines (experimental data, Tab. 1 and Tab. 2 in the text) are used in the calculations. S corresponds to the scaling factor (determined using the 1630 cm$^{-1}$ band of $Bz^+$, cf. Fig. S9).

## 2.6 Influence of the Ne tag for Tr$^+$

Table S4: Shift of the Tr$^+$ band around 650 cm$^{-1}$ as a function of the Tr$^+ \cdot$ Ne configuration for different calculation methods. The shift predicted by the best levels of theory used to describe the systems studied in this work (wB97XD/cc-pvtz and MP2/cc-pvtz) is just below the resolution of our experiment.

| Method | Tr$^+$ | +Ne (top) | +Ne (plane) | $\Delta$ Ne (plane − top) |
|---|---|---|---|---|
| B3LYP/6-31G(d,p) | 663.6 | 660.0 | 671.2 | 11.2 |
| B3LYP/cc-pvtz | 664.8 | 662.9 | 666.6 | 3.7 |
| wB97XD/cc-pvdz | 665.9 | 663.1 | 671.0 | 7.9 |
| wB97XD/cc-pvtz | 648.6 | 646.3 | 649.8 | 3.5 |
| MP2/cc-pvtz | 660.6 | 658.5 | 661.6 | 3.1 |
| B2PLYP/cc-pvtz | 665.8 | 663.8 | 667.3 | 3.5 |

**Note:** All frequencies in cm$^{-1}$. Frequencies are not scaled. In Ne IR-PD experiment, the distance between the two observed bands is 22 cm$^{-1}$.

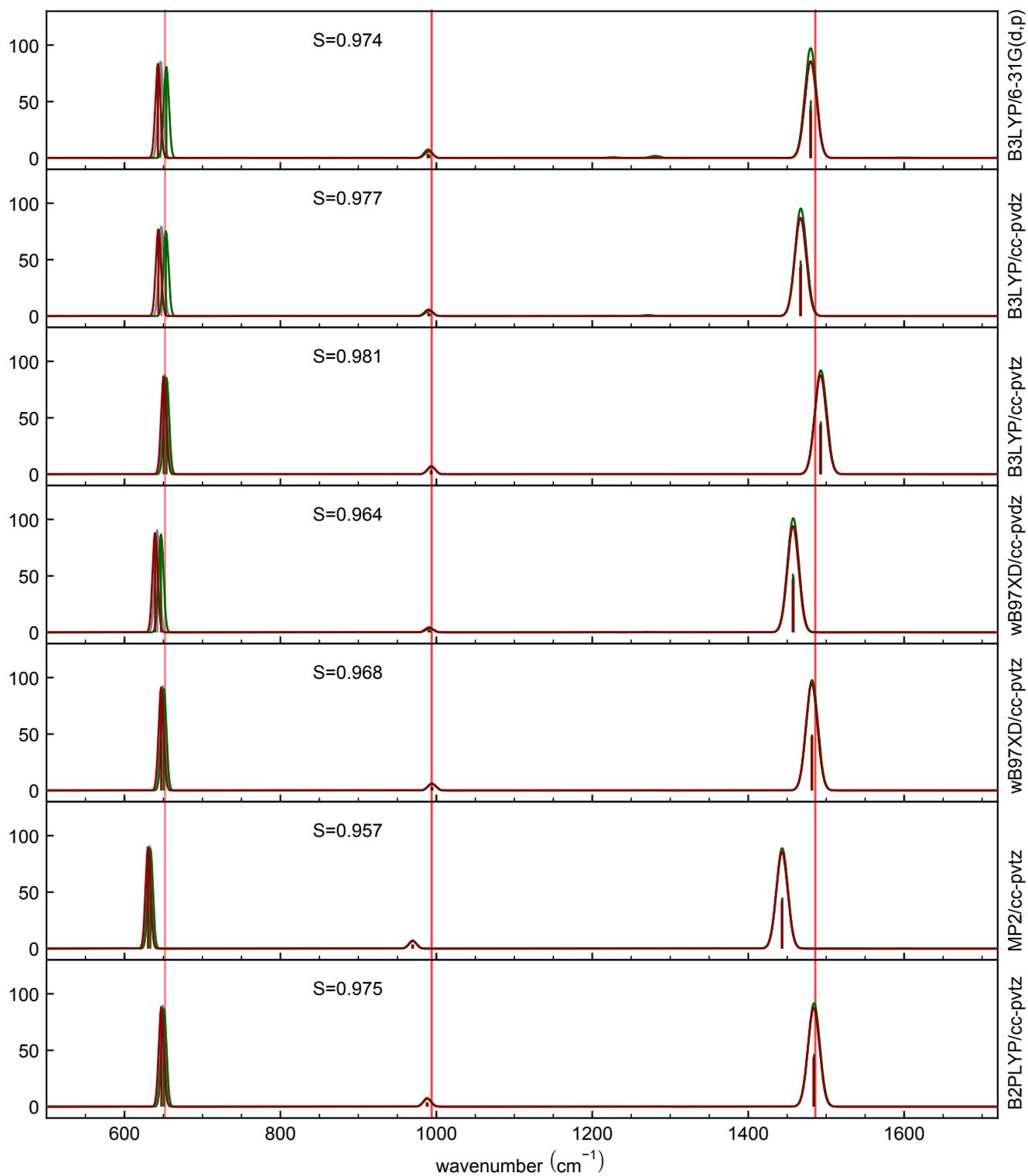

Figure S12: Spectra corresponding to different positions of the Ne atom attached to Tr$^+$. S corresponds to the scaling factor (determined using the 1630 cm$^{-1}$ band of Bz$^+$, cf. Fig. S9). Color code: grey – bare ion, dark red – Ne on top of the molecular plane (cf. Figs. S4), dark green – Ne in the molecular plane (cf. Figs. S5).

## 2.7 Influence of the Ne tag for Bz$^+$

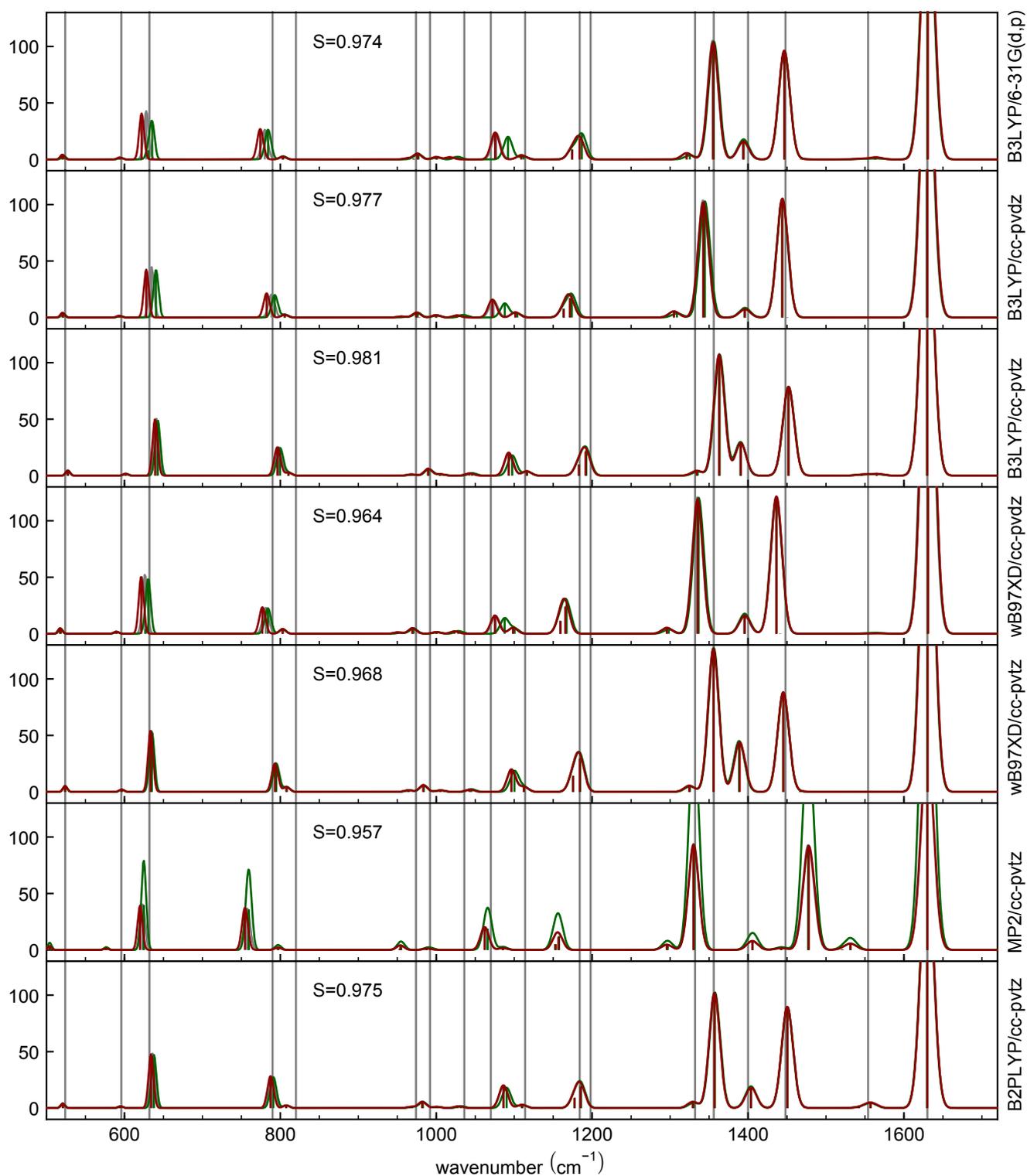

Figure S13: Spectra corresponding to different positions of the Ne atom attached to Bz$^+$. S corresponds to the scaling factor (determined using the 1630 cm$^{-1}$ band of Bz$^+$, cf. Fig. S9). Color code: grey – bare ion, dark red – Ne on top of the molecular plane (cf. Figs. S6), dark green – Ne in the molecular plane (cf. Figs. S7).

## 2.8 Combination band of Tr$^+$

Table S5: Comparison of position of the Tr$^+$ and Tr$^+ \cdot$Ne out of plane CH bend around 650 cm$^{-1}$ with possible combination bands.

|  | B3LYP/6-31G(d,p) S=0.974 | | | wB97XD/cc-pvtz S=0.968 | | | MP2/cc-pvtz S=0.957 | | |
| --- | --- | --- | --- | --- | --- | --- | --- | --- | --- |
|  | Tr$^+$ | +Ne (T) | +Ne (P) | Tr$^+$ | +Ne (T) | +Ne (P) | Tr$^+$ | +Ne (T) | +Ne (P) |
| $\omega_{\text{Ne 0}}$ |  | 42.3 | 40.3 |  | 20.6 | 19.6 |  | 28.4 | 15.6 |
| $\omega_{\text{Ne 1}}$ |  | 42.7 | 59.2 |  | 26.5 | 27.6 |  | 28.4 | 30.9 |
| $\omega_{\text{Ne 2}}$ |  | 96.0 | 106.7 |  | 42.3 | 50.4 |  | 56.7 | 47.8 |
| $\omega_0$ | 218.4 | 211.6 | 220.1 | 213.5 | 211.0 | 213.8 | 206.0 | 203.6 | 206.4 |
| $\omega_1$ | 218.9 | 212.3 | 224.9 | 213.7 | 212.4 | 214.9 | 206.0 | 203.6 | 207.5 |
| $\omega_2$ | 427.2 | 427.1 | 428.2 | 426.7 | 426.2 | 427.2 | 407.8 | 407.2 | 407.9 |
| $\omega_3$ | 427.7 | 427.7 | 428.2 | 427.1 | 426.7 | 427.4 | 407.8 | 407.2 | 408.1 |
| $\omega_4$ | 549.4 | 541.2 | 551.7 | 552.1 | 549.7 | 552.7 | 531.0 | 528.1 | 531.9 |
| $\omega_5$ | 549.6 | 541.6 | 552.7 | 552.3 | 550.0 | 553.1 | 531.0 | 528.1 | 532.2 |
| $\omega_6$ | 646.4 | 642.9 | 653.7 | 649.3 | 647.0 | 650.5 | 632.2 | 630.2 | 633.2 |
| $\omega_{\text{Ne 2}} + \omega_4$ |  | 637.2 | 658.3 |  | 591.9 | 603.1 |  | 584.8 | 579.7 |
| $\omega_0 + \omega_2$ | 645.6 | 638.7 | 648.3 | 640.1 | 637.1 | 641.0 | 613.9 | 610.8 | 614.3 |

**Note:** All frequencies in cm$^{-1}$. All modes except $\omega_6$ ($I \sim 80$ km$\cdot$mol$^{-1}$) and $\omega_{\text{Ne 2}}$ ($I \sim 3$ km$\cdot$mol$^{-1}$) have zero intensities. (T) – Ne on the of the molecular plane, (P) – Ne in the molecular plane.

## 2.9 Excited electronic state of Tr$^+$

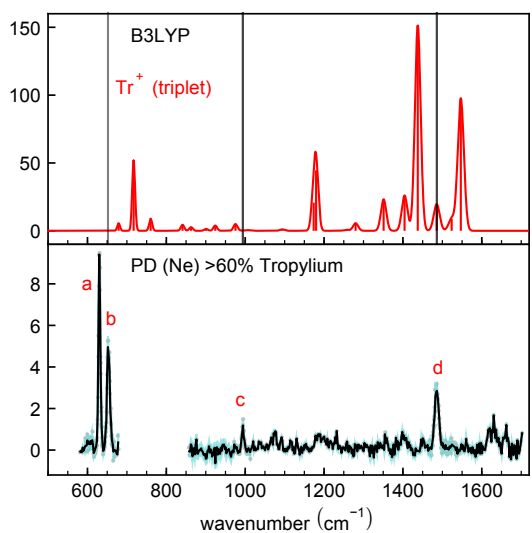

Figure S14: Calculated frequencies of the triplet state of Tr$^+$ at the B3LYP/6-31G(d,p) level of theory (scaling factor 0.974) compared to the Ne IR-PD spectrum of Tr$^+$.

# 3  Visualisation of vibrational modes

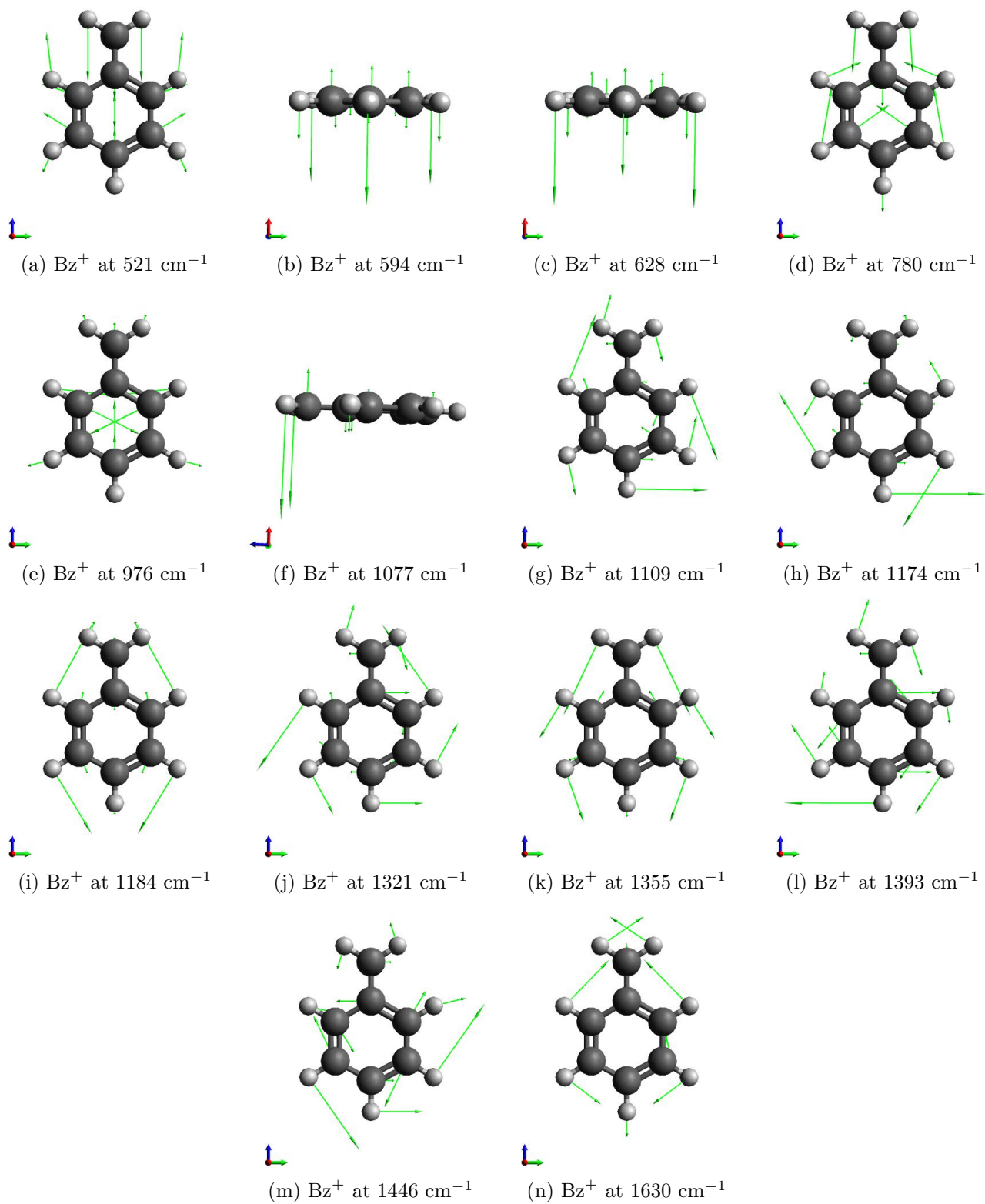

(a) Bz$^+$ at 521 cm$^{-1}$
(b) Bz$^+$ at 594 cm$^{-1}$
(c) Bz$^+$ at 628 cm$^{-1}$
(d) Bz$^+$ at 780 cm$^{-1}$
(e) Bz$^+$ at 976 cm$^{-1}$
(f) Bz$^+$ at 1077 cm$^{-1}$
(g) Bz$^+$ at 1109 cm$^{-1}$
(h) Bz$^+$ at 1174 cm$^{-1}$
(i) Bz$^+$ at 1184 cm$^{-1}$
(j) Bz$^+$ at 1321 cm$^{-1}$
(k) Bz$^+$ at 1355 cm$^{-1}$
(l) Bz$^+$ at 1393 cm$^{-1}$
(m) Bz$^+$ at 1446 cm$^{-1}$
(n) Bz$^+$ at 1630 cm$^{-1}$

Figure S15: Harmonic vibrational modes of Bz$^+$ calculated at the B3LYP/6-31G(d,p) level of theory, scaling factor 0.974.

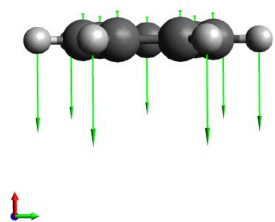

(a) Tr$^+$ at 646 cm$^{-1}$

Figure S16: Harmonic vibrational modes of Tr$^+$ calculated at the B3LYP/6-31G(d,p) level of theory, scaling factor 0.974. Out of plane CH bending.

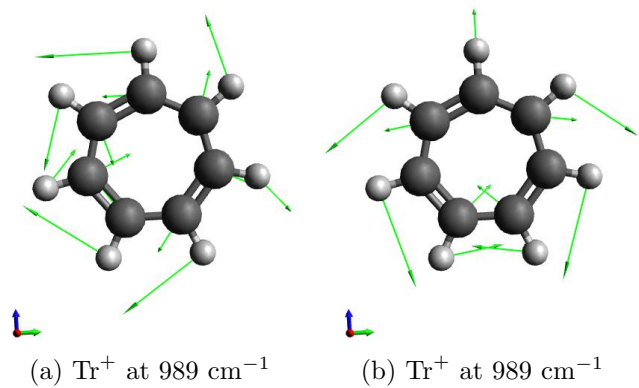

(a) Tr$^+$ at 989 cm$^{-1}$  (b) Tr$^+$ at 989 cm$^{-1}$

Figure S17: Harmonic vibrational modes of Tr$^+$ calculated at the B3LYP/6-31G(d,p) level of theory, scaling factor 0.974. In plane CH bending mode.

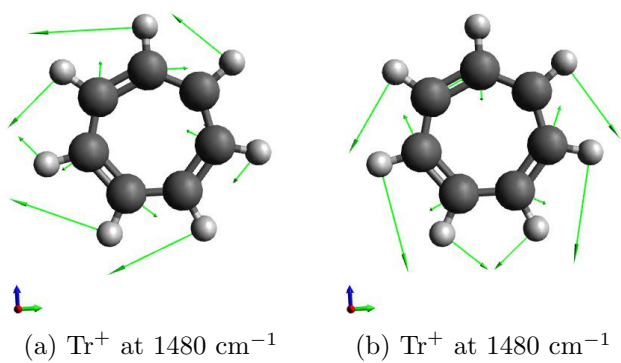

(a) Tr$^+$ at 1480 cm$^{-1}$  (b) Tr$^+$ at 1480 cm$^{-1}$

Figure S18: Harmonic vibrational modes of Tr$^+$ calculated at the B3LYP/6-31G(d,p) level of theory, scaling factor 0.974. In plane CC stretching mode.



\end{document}